\documentclass[superscriptaddress,twocolumn,prl,noeprint]{revtex4-1}

\usepackage{amsfonts}
\usepackage{amsmath}    % for subequations
\usepackage[normalem]{ulem}
\usepackage{bm} %% bold math package for making greek letters bold
\usepackage{graphicx}
\usepackage{titlesec}
\usepackage{tabularx}
\usepackage{color}
\usepackage{bbold}
\usepackage{upgreek,ulem,textcomp}
\usepackage{bbold}

\usepackage[colorlinks=true,linkcolor = blue,citecolor = blue,urlcolor= blue]{hyperref}
\usepackage{hypernat}

\usepackage{amsfonts}
\usepackage{amsmath}    % for subequations
\usepackage[normalem]{ulem}
\usepackage{bm}
\usepackage{graphicx}
\usepackage{titlesec}
\usepackage{tabularx}
\usepackage{xcolor}
\usepackage{bbold}
\usepackage{upgreek,ulem,textcomp}
\usepackage{bbold}
\usepackage{comment}
\usepackage[british]{babel}
\usepackage{csquotes}
\usepackage{etoolbox}

\newcommand{\ket}[1]{\vert#1\rangle}

\newcommand{\rmnum}[1]{\romannumeral #1}
\newcommand{\Rmnum}[1]{\expandafter\@slowromancap\romannumeral #1@} %% Roman numerals

\begin{abstract}
Superradiance, characterized by the collective, coherent emission of light from an excited ensemble of emitters, generates photonic signals on timescales faster than the natural lifetime of an individual atom. The rapid exchange of coherence between atomic emitters and photonic fields in the superradiant regime enables a fast, broadband quantum memory. We demonstrate this superradiance memory mechanism in an ensemble of cold rubidium atoms and verify that this protocol is suitable for pulses on timescales shorter than the atoms' natural lifetime.  Our simulations show that the superradiance memory protocol yields the highest bandwidth storage among protocols in the same system. These high-bandwidth quantum memories provide unique opportunities for fast processing of optical and microwave photonic signals, with applications in large-scale quantum communication and quantum computing technologies.

\end{abstract}

\begin{document}
% \title{Superradiance effect for broadband light storage in cold atoms}
\title{Superradiance-mediated photon storage for broadband quantum memory}

\author{Anindya Rastogi}
\affiliation{Department of Physics, University of Alberta, Edmonton AB T6G 2E1, Canada}
\author{Erhan Saglamyurek}
\affiliation{Department of Physics, University of Alberta, Edmonton AB T6G 2E1, Canada}
\affiliation{Department of Physics and Astronomy, University of Calgary, Calgary, Alberta T2N 1N4, Canada}
\author{Taras Hrushevskyi}
\affiliation{Department of Physics, University of Alberta, Edmonton AB T6G 2E1, Canada}
\author{Lindsay J. LeBlanc}
\email{Corresponding authors: lindsay.leblanc@ualberta.ca, rastogi1@ualberta.ca }
\affiliation{Department of Physics, University of Alberta, Edmonton AB T6G 2E1, Canada}

\maketitle 
Photonic emission from a collection of identical excited atoms under superradiant conditions is very different from that of a single-atom in free-space~\cite{PhysRev.93.99}. Comprehensive studies of superradiance~\cite{andreev1980collective,gross1982superradiance,cong2016dicke} have led to the observation of various quantum optics phenomena like quantum beats~\cite{han2021observation}, collective Lamb shifts~\cite{scully2009collective,svidzinsky2009evolution}, and novel cavity QED~\cite{doi:10.1126/science.1176695}. Initial studies focused on dense ensembles with sizes smaller than the excitation wavelength, whereas recent ones show that superradiance is also observable in large and dilute systems~\cite{araujo2016superradiance,PhysRevLett.117.073003}. Large atomic ensembles, providing a high degree of experimental control, are especially amenable to superradiant effects: when an ensemble with moderate optical density ($d > 1$) is coherently excited by a temporally short (broadband) and weak probe, a subsequent radiation-burst is emitted along the forward direction of the incident pulse~\cite{kwong2014cooperative,kwong2015cooperative,chalony2011coherent,dudovich2002coherent,costanzo2016zero}. The characteristic decay time of this emission is inversely proportional to the medium's optical depth and, for $d \gg 1$, can be much shorter than the spontaneous-emission lifetime of a single atom~\cite{scully2006directed,svidzinsky2008dynamical,bienaime2011atom,sutherland2016coherent,bromley2016collective,araujo2016superradiance,PhysRevLett.117.073003,han2021observation}. This optical-depth dependence of decay time is the hallmark of superradiant emission, different from the free-induction-decay (FID) of atomic dipoles among inhomogeneously broadened emitters~\cite{vivoli2013high,Walther2009}.

Optical quantum memories~\cite{lvovsky2009optical,hammerer2010quantum, heshami2016quantum} can take advantage of this rapid superradiant emission for broadband operation~\cite{kalachev2007quantum,gorshkov2007photon,vivoli2013high}. Here, we demonstrate a spin-wave memory based on superradiance~\cite{Footnote1,cong2016dicke}  in a cloud of laser-cooled $^{87}$Rb atoms featuring a homogeneously-broadened optical transition.  
\textcolor{black}{The superradiance (SR) mediated  memory requires an initial excitation of the polarization coherence, followed by a second (fast) storage step, realizing a distinct memory mechanism that is inherently broadband and, in our experiments, is demonstrated without contribution from physical processes associated with other memory protocols.} We find that the SR memory offers the most relaxed optical depth requirement for broadband signals in systems where the absorption linewidths are much narrower than the signal bandwidth to be stored.
The inherently fast storage capability of this memory paves the way for bandwidth compatibility with conventional single-photon sources and high-speed quantum networks~\cite{sangouard2011quantum,kimble2008quantum}.

\begin{figure}[t!]
\begin{center}
\includegraphics[width = 85 mm]{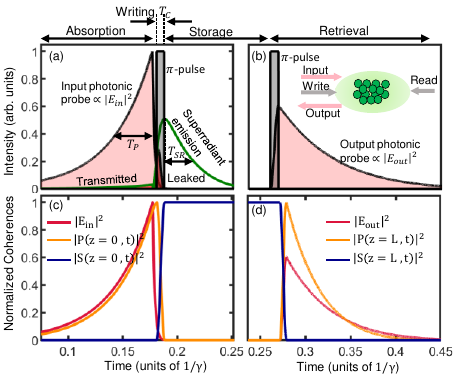}
\caption{Superradiance (SR) memory protocol. (a) A short input probe (pink) with duration $T_{\rm P} \ll 1/\gamma$ is absorbed by an ensemble with $d \gg 1$, and, absent the memory control, superradiantly reemits (green) with decay time $T_{\rm SR} \ll 1/\Gamma$. When a $\pi$-control (grey) with a duration $T_{\rm C} \ll {T}_{\rm SR}$ is applied, it suppresses  superradiant emission and induces photonic storage. (b) Re-applying the $\pi$-pulse (grey) results in emission of the stored probe. (c) Time evolution of (normalized) photonic (pink), polarization (orange), and spin-wave (blue) coherences at the input face of the medium, showing probe storage. (d) Coherences upon readout, in backward recall. Simulation parameters \big[$T_{\rm P},d,T_{\rm C},B,\Omega_{\rm C}, \eta$\big] are \big[$0.037/\gamma, 50, 0.0074/\gamma, 13.3\Gamma/2\pi, 207\Gamma, 0.79$\big].} 
\label{fig:protocol}
\end{center}
\end{figure}

To understand the SR memory, consider an $N$-atom ensemble  whose energy levels form a $\Lambda$-configuration, where two ground-state spin levels $\ket{g}$ and $\ket{s}$ are optically coupled to an excited level $\ket{e}$. We assume that all atoms initially populate $\ket{g}$, and that both $\ket{e} \leftrightarrow \{\ket{g},\ket{s}\}$ transitions are homogeneously-broadened, with Lorentzian lineshapes of characteristic width $\Gamma$ and optical decoherence rates $\gamma = \Gamma/2$. A weak probe field (the optical signal to be stored) is resonant with the $\ket{g} \leftrightarrow \ket{e}$ transition, and a strong control field (to initiate probe storage or recall) with Rabi frequency $\Omega_{\rm C}(t)$ drives the $\ket{e} \leftrightarrow \ket{s}$ transition.~Resonant interactions between the optical fields and ensemble are described by the coupled Maxwell-Bloch equations \textcolor{black}{in terms of coherences in the photonic ($\hat{E}$), polarization ($\hat{P}$), and spin-wave ($\hat{S}$) modes}~\cite{gorshkov2007photon,gorshkov2007universal,sangouard2007analysis,nunn2007mapping,supplementary}. 
For the input probe, we consider an exponentially rising temporal profile of the form $I_{\rm P}(t) = I_{\rm 0}e^{(t-\tau)/T_{\rm P}} u(\tau - t)$ [$\tau$ is the abrupt switch-off time]; characteristic duration $T_{\rm P} \ll 1/\gamma$ (Fig~\ref{fig:protocol}); and bandwidth $B = 0.157/T_{\rm P}$~\cite{supplementary}, such that  memory operation is broadband with $2\pi B \gg \Gamma$.

The SR memory protocol (Fig.~\ref{fig:protocol}) proceeds via three stages: absorption, writing, and retrieval, which follow the general principles of ``fast’’ lambda-type storage and readout~\cite{gorshkov2007photon,gorshkov2007universal}, and can also be extended to the ``ladder’’-type atomic configurations~\cite{finkelstein2018fast,Kaczmarek2018high}. During absorption, the short incident probe builds up  polarization coherence $|P(z,t)|^{2}$  (Fig.~\ref{fig:protocol}c) across the entire ensemble~\cite{supplementary} on a timescale much shorter than the excited-state decoherence, which can therefore be neglected ($e^{-\gamma T_{\rm P} }\approx$ 1). Atomic polarization is maximized at the conclusion of the input pulse, leading to a subsequent superradiant reemission (Fig.~\ref{fig:protocol}a). Regardless of the line broadening mechanism (homogeneous or inhomogeneous), such reemission occurs naturally whenever a broadband pulse is absorbed by a spectral feature with linewidth narrower than the bandwidth of the incident pulse~\cite{vivoli2013high}. This absorption regime differs from that of photon-echo and FID processes, where the probe bandwidths must be smaller than or comparable to the width of the inhomogeneously broadened emitters~\cite{tittel2010photon}.   

In the writing stage, storage is achieved by converting the built-up polarization into a collective spin excitation via a control field (Fig.~\ref{fig:protocol}a) before superradiant emission proceeds. For efficient $\hat{P} \rightarrow \hat{S}$ mapping, (\rmnum{1}) the control duration ($T_{\rm C}$) must be much shorter than the superradiant decay time \big($T_{\rm C} \ll T_{\rm SR}$\big) ensuring minimum leakage, and (\rmnum{2}) the control pulse-area must be $\pi$, ensuring maximum transfer~\cite{gorshkov2007photon,vivoli2013high,ho2018optimal,supplementary}. \textcolor{black}{Like all spin-wave memories,} the initial coherence remains stored up to a time limited by the spin-wave decoherence. Finally, retrieval is implemented by re-applying the $\pi$-pulse to map the coherence from $\hat{S} \rightarrow \hat{P} \rightarrow \hat{E}$, recovering the probe into the output photonic mode (Fig.~\ref{fig:protocol}b,d). \textcolor{black}{Bidirectional emission, as observed in some superradiance experiments~\cite{Sadler2007,Hilliard2008,Uys2007}, is not a factor in this memory system. The phase pattern of excited dipoles is imposed by the interference between probe and control fields, and phase matching conditions ensure that the retrieved probe is emitted into the desired mode.}

\begin{figure*}[tb!] %% Use asterisk command to allow a figure to span the two columns.
% \begin{center}
\includegraphics[width = 170 mm] {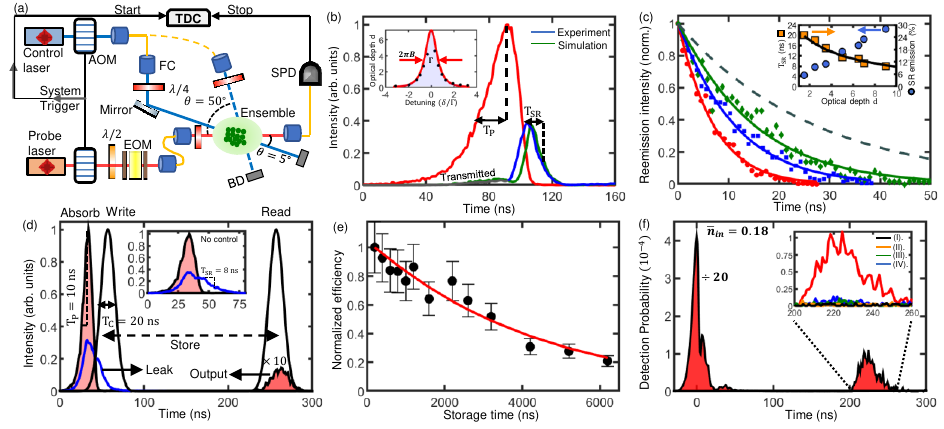}
\caption{Demonstration of SR memory in cold $^{87}$Rb atoms. {(a)} Schematic of the setup. \textcolor{black}{FC: fiber coupler, AOM: acousto-optic-modulator, EOM: electro-optic-modulator, SPD: single-photon-detector, TDC: time-to-digital converter, BD: beam dump.} {(b)} Superradiant emission following an exponential probe. Inset: probe spectral profile (solid red) with bandwidth $B = 1.23\Gamma/2\pi$ for linewidth $\Gamma = 2\pi \times$ 6 MHz (shaded blue). {(c)} Decay of the emission intensity (normalized to its own maximum) for optical depth 9 (red circles), 5 (blue squares), and 3.5 (green diamonds), with solid lines fit to the data. Black dashed line shows spontaneous-emission decay (independent of $d$). Inset: superradiant decay time $T_{\rm SR}$ and emission efficiency versus optical depth. Solid black line is fit of  measured $T_{\rm SR}$ values using Eq.~(\ref{eqn:opt_dur}), \textcolor{black}{which matches the probe duration for optimal storage via the SR protocol}. {(d)} Storage and forward retrieval of a $T_{\rm P}$ = 10~ns, $B$ = 12.7 MHz probe containing $\bar{n}_{\rm in} \gg 1$. Inset: SR emission without control. {(e)} Variation of memory efficiency with storage time~\cite{supplementary}. {(f)} Storage and recall at the single-photon level with $\bar{n}_{\rm in}$= 0.18. Inset: memory-retrieved signal intensity relative to noise measured in four configurations (other colors)~\cite{supplementary}.}
\label{fig:experimental results}
% \end{center}
\end{figure*}

\textcolor{black}{To experimentally demonstrate the SR memory, we use an ensemble of $N_{\rm A} \approx 2\times 10^{8}$ laser-cooled $^{87}$Rb atoms. With $1/e^2$ Gaussian diameters of 2.5~mm, 3.6~mm, and  4~mm, the cloud has  volume density  $7 \times 10^9$ cm$^{-3}$ and peak optical depth  $d = 9$ (along the longest direction)}. We form the $\Lambda$-configuration on the ``D2'' line, with levels $\ket{g} \equiv \ket{5S_{1/2},  F = 1}$, $\ket{s} \equiv  \ket{5S_{1/2}, \:  F = 2}$, and $\ket{e} \equiv \ket{5P_{3/2}, \: F' = 2}$. The probe and control fields, resonant with the $\ket{g} \leftrightarrow \ket{e}$ and $\ket{s} \leftrightarrow \ket{e}$ transitions, are derived from two independent, phase-locked lasers, and temporally shaped via electro-optic (EOM) and acousto-optic (AOM) modulators, respectively (Fig.~\ref{fig:experimental results}a). \textcolor{black}{Depending on   mean photon number per probe pulse $(\bar{n}_{\rm in})$, the control field is oriented either at $\theta = 5^{\circ}$ (high $\bar{n}_{\rm in} \sim 10^3$) or $\theta = 50^\circ$ (single-photon-level probe with $\bar{n}_{\rm in} < 1$, where the large angle permits spatial filtering of scattered control noise~\cite{saglamyurek2021storing}) relative to the probe axis}. The output probe is detected along the forward direction using a single-photon detector (SPD) and a time-to-digital counter (TDC). \textcolor{black}{An experimental sequence begins with ensemble preparation, followed by $1000$ measurement-and-detection events performed over 1~ms.}

To begin, we characterize the superradiant behaviour of our cold Rb ensemble by sending short \big($T_{\rm P}$ = 20 ns $ < 1/\gamma$ = 54 ns\big) probe pulses with increasing-exponential profiles and sharp switch-off (Fig.~\ref{fig:experimental results}b). The probe is almost completely absorbed up to the switch-off time, and subsequently reemitted along the forward direction, as superradiance with characteristic time $T_{\rm SR} = 8$~ns. 
The observed superradiant decay time is shorter than both the input duration and the spontaneous emission lifetime~\cite{araujo2016superradiance,PhysRevLett.117.073003,han2021observation,PhysRevLett.109.263601,PhysRevLett.116.233601,PhysRevA.97.053816}, by factors of 2.5 and 3.5, respectively. 
A linewidth ($\Gamma$) measurement  of the probe transition \textcolor{black}{confirms no inhomogeneous broadening~\cite{Maki1989influence,Malcuit1987transition}, which together with the probe bandwidth $2\pi B > \Gamma$, shows the conditions for superradiance are satisfied (Fig.~\ref{fig:experimental results}b, inset).}  

Next, we verify the superradiant nature of  reemission by measuring its decay time as a function of optical depth (Fig.~\ref{fig:experimental results}c). We control the optical depth between $d = 1.5$ to 9 by varying the atomic cloud's time-of-flight spatial expansion before measurement. For each $d$, we extract the $1/e$ decay time  ($T_{\rm SR}$), shown in Fig.~\ref{fig:experimental results}c inset, with times from $T_{\rm SR} = \big(8.0 \pm 0.1$\big)~ns to $T_{\rm SR} = \big(18.8 \pm 0.8$\big)~ns, confirming emission's superradiant behaviour: \textcolor{black}{the enhancement of the decay rate with respect to optical depth is the characteristic signature of superradiance~\cite{araujo2016superradiance,PhysRevLett.117.073003,han2021observation,PhysRevLett.109.263601,PhysRevLett.116.233601,PhysRevA.97.053816}}. This inset also shows SR emission efficiency: the higher the $d$, the greater the energy contained within the emitted signal and the shorter the emission timescale, suggesting that high memory efficiency requires both large optical depth (leading to an efficient polarization build-up) and fast writing. 

%%% Modified experimental description to include new plots
We demonstrate the \textcolor{black}{distinct} operation of the SR spin-wave memory by storing an exponentially rising 10-ns probe (containing $\bar{n}_{\rm in} \approx 1.5\times10^3$ photons), and retrieving it along the forward direction, using Gaussian write and read control pulses of full-width-half-maximum (FWHM) duration $T_{\rm C} = 20$~ns \textcolor{black}{and peak power 8~mW} (Fig.~\ref{fig:experimental results}d). \textcolor{black}{The write control is applied 25~ns after the probe, ensuring the storage process is activated only by the SR memory mechanism~\cite{supplementary}}. 
By changing the time-interval between write and read stages, we observe an exponential decay of the memory efficiency yielding a $1/e$ memory-lifetime of ($4.2 \pm 0.3$)~$\mu$s (Fig.~\ref{fig:experimental results}e), limited by the spin-wave decoherence induced by ambient magnetic fields. At 200 ns storage-time, the efficiency is 3\%, significantly lower than the maximum achievable SR memory efficiency of 31$\%$, calculated for $d = 9$ and forward recall. Achieving this efficiency  requires optimization of both probe and control fields. 

An optimized probe is characterized by an exponentially rising temporal shape with a duration that matches the superradiant decay of our system, which makes it broadband, since $B \propto 1/T_{\rm P} \approx 1/T_{\rm SR} > \Gamma$. In our demonstration, both the probe shape and duration \big[$T_{\rm P}=10$ ns $\approx T_{\rm SR}=8$ ns\big] adequately fulfill these optimal conditions~\cite{supplementary}, but the efficiency is limited by non-optimized control: even at our maximum power, the shortest $\pi$-pulse we can generate exceeds $T_{\rm SR}$, violating the condition $T_{\rm C} \ll T_{\rm SR}$. As a result, a significant fraction of the polarization coherence $\hat{P}$ is lost via superradiant emission before it can be mapped to the spin-wave $\hat{S}$. This loss could be eliminated by using a nanosecond-long control, which would require two orders of magnitude more intensity for the $\pi$-pulse. 
\textcolor{black}{Still, in the absence of such intensity, the fast-writing process is evident in these experiments. First, the temporal separation between probe and write-control ensures the storage mechanisms from other protocols are suppressed~\cite{supplementary};  second, when the system is prepared at a lower optical depth ($d = 6$) and consequently longer $T_{\rm SR}$ (11~ns), the maximum power acts over this longer $T_{\rm SR}$ and transfers a greater proportion of $\hat{P}$ to $\hat{S}$, increasing the memory efficiency (from 3\% to 5\%)~\cite{supplementary}.~This counterintuitive increase in efficiency for lower $d$ under non-optimal conditions is characteristic of the SR mechanism, and would not be found in alternate memory mechanisms. 
Additionally, we find that under similar conditions, the efficiency for longer probe pulses (lower bandwidths) decreases, indicating that  SR memory operates best for shorter pulses, up to where $T_{\rm P}$ is comparable to $ T_{\rm SR}$~\cite{supplementary}.}

%%%%%%%%%%%%%  Added measurement details on SR memory operation with probe pulses at the single photon level  
\textcolor{black}{Next, we investigate the compatibility of SR memory with quantum signals by operating with probe pulses at the single-photon level. In particular, we measure noise from extraneously added photons during the memory operation and use this to predict fidelity for quantum state storage. We measure an unconditional noise probability of $(2.1 \pm 0.2)\times 10^{-4}$~\cite{supplementary}, which corresponds to a stored-and-recalled signal-to-noise ratio (SNR) of $ 12.4 \pm 1.2$ for a 10-ns probe with $\bar{n}_{\rm in} =0.18$ (Fig.~\ref{fig:experimental results}f). If quantum states were encoded in these pulses, this SNR indicates a maximum quantum storage fidelity of $F= 1-1/\text{SNR}= 0.92 \pm 0.09$. Experimentally, memory efficiency remains constant across different values of $\bar{n}_{\rm in}$, yielding a linear increase in SNR~\cite{supplementary}. Importantly, SNR is limited only by  noise from scattered control beams and not by any physical process linked to memory operation, suggesting that the SR protocol will reliably store quantum signals, such as those encoded in space, polarization, or time bins}.

%%%%%%%%%%%%%%%%%%%%%%%%%%%%%%%%%%%%%%%%%%%%%%%%%%%%%%%%%%%%%%%%%%%%%%%%%%%%%%%%%%%%%%%%%%%%%
%%%%%%% Optimality section
With these proof-of-concept results at hand, we consider future implementations by examining  optimality conditions for the SR protocol. Independent of any memory approach, the ``optimality'' criterion dictates that the maximum achievable memory efficiency ($\eta_{\rm opt}$) is universal and depends only on the medium's optical depth~\cite{gorshkov2007photon,gorshkov2007universal}, and can be satisfied by implementing protocol-specific  optimizations of both the probe and the control, under the backward-recall configuration.

Probe optimization in the SR protocol relies on both maximizing the polarization build-up and ensuring its proper spatial-distribution \big[$P(z)$\big] in the absorption stage, which leads to an optimal spin-wave mode in the subsequent writing and readout stages. This is achieved by an input temporal profile that matches the time reversed replica of the superradiantly emitted pulse, both in shape and duration~\cite{Walther2009,kalachev2007quantum,PhysRevA.80.012317,stobinska2009perfect,dao2012preparation}. For a Lorentzian spectral feature (as in our cold Rb system, Fig.~\ref{fig:experimental results}b, inset), the optimal shape of this probe is an exponentially-rising envelope with its duration given by~\cite{vivoli2013high,ho2018optimal}
\begin{equation}
    T_{\rm P}^{\rm opt} \approx \frac{1}{\Gamma}\left(\frac{1}{1 + {d}/{4}}\right), \label{eqn:opt_dur}
\end{equation}
which corresponds to the system's $T_{\rm SR}$ emission time, as verified numerically and experimentally (solid fit in Fig.~\ref{fig:experimental results}c, inset).  \textcolor{black}{For optical depths $d > 1$, the pulse time is shorter than spontaneous emission lifetime, and thus the probe bandwidth exceeds $\Gamma$, rendering this protocol inherently broadband.}

Control optimization requires write-and-readout $\pi$-pulses with durations $T_{\rm C}$ much shorter than $T_{\rm SR}$. For square control pulses, numerical simulations verify that $T_{\rm C}^{\rm opt} \approx  T_{\rm SR}/10 \approx  T_{\rm P}^{\rm opt}/10$ is sufficient for optimal efficiency, in turn requiring $\Omega_{\rm C} \gg \gamma(1 + d)$ for fast writing and retrieval~\cite{gorshkov2007photon,gorshkov2007universal}. For the Gaussian controls used in our experiments (with $d = 9$), this translates to a peak Rabi frequency $\Omega_{\rm C}^{\rm opt} = 130\gamma$, well above our experimental $\Omega_{\rm C}^{\rm exp}= 4.8 \gamma$, accounting for the memory inefficiency. 

Under optimized probe-and-control conditions, we investigate the optical-depth dependence of probe-duration (bandwidth) in terms of the adiabaticity parameter \big($T_{\rm P}^{\rm opt} d \gamma$\big).  Using Eq.~(\ref{eqn:opt_dur}) and for $d \gg 1$,  $T_{\rm P}^{\rm opt} d \gamma \leq 2$, which represents the ``fast'' operating regime opposite to that of adiabatic memories, which instead are characterized by $T_{\rm P}^{\rm opt} d \gamma \gg 1$~\cite{gorshkov2007photon,gorshkov2007universal,nunn2007mapping}. Clearly, SR memory falls into the class of non-adiabatic memories~\cite{PhysRevA.76.033806,tittel2010photon}, which are inherently suitable for optimal storage of broadband signals \big($2 \pi B > \Gamma$\big). 
\textcolor{black}{We note, however, that in homogeneously-broadened media, the bandwidth of the SR control fields is ultimately limited by the spacing between ground levels. For example, storage of a one-nanosecond probe would require control fields with GHz spectral width, comparable to the ground-state hyperfine splitting of alkali atoms. Such broad control fields may off-resonantly excite atoms in $\ket{g}$, creating spurious spin-waves that can add photonic noise at the memory output via four-wave mixing (FWM)~\cite{Phillips2011,Michelberger2015,heshami2016quantum,Lauk2013,Thomas2019}.}

\begin{figure}
\begin{center}
\includegraphics[width = 85 mm] {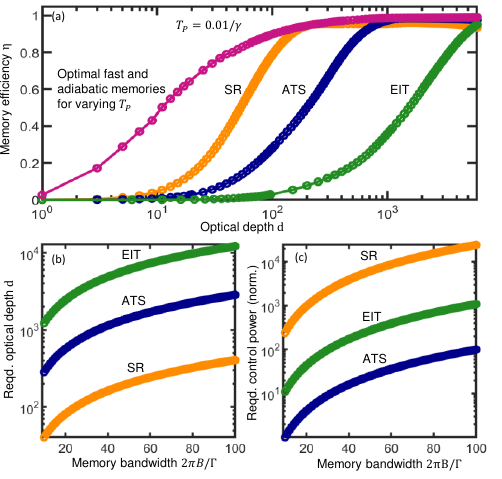}
\caption{Theoretical analysis of SR memory versus ATS and EIT memories for storage of broadband pulses with optimized parameters \big[$d,T_{\rm P},B,\Omega_{\rm C}(t)$\big] in backward recall. ATS and EIT protocols are optimized using~\cite{saglamyurek2018coherent,PhysRevA.100.012314}. {(a)} Numerically calculated memory efficiency vs.\ optical depth for an exponential probe with $T_{\rm P} = 0.01/\gamma$ ($B = 49.4\Gamma/2\pi$), showing optimal operation at ($d =$ 200, $\eta = 0.93$) ; ($d =$ 1400, $\eta = 0.96$) ; and ($d =$ 6000, $\eta = 0.93$) for the SR, ATS, and EIT protocols respectively. Universal optimal efficiency $\eta_{\rm opt}$ is obtained via optimal implementation of any protocol. {(b,c)} Optical depth and control power requirements versus memory bandwidth, where near-optimal operation is maintained ($\eta \ge$ 0.9). Control strength is normalized with respect to the ATS power required for $B =$ 10$\Gamma/2\pi$.} \label{fig:performance}
\end{center}
\end{figure}

Finally, we make a performance comparison (in terms of efficiency, optical depth, bandwidth, and control power) between SR and other memory approaches, including Autler-Townes splitting (ATS)~\cite{saglamyurek2018coherent,PhysRevA.100.012314,saglamyurek2019single,saglamyurek2021storing} and  electromagnetically induced transparency (EIT)~\cite{PhysRevLett.84.5094,PhysRevA.65.022314,hsiao2018highly}, which are examples of non-adiabatic and adiabatic protocols, respectively. We base this comparison on typical adiabaticity parameters, corresponding to $T_{\rm P} d \gamma = 2$ (SR); $T_{\rm P} d \gamma = 14$ (ATS); and $T_{\rm P} d \gamma = 60$ (EIT), for a broadband, exponential probe with $T_{\rm P} \ll 1/\gamma$. For a given bandwidth (probe duration), the optical depth required for optimal SR efficiency is 7 times lower than in an ATS memory and 30 times lower than in an EIT memory (Fig.~\ref{fig:performance}a). Equivalently, for a given optical depth, the bandwidth that can be optimally stored using the SR protocol is higher than the corresponding bandwidth in the ATS and EIT protocols by the same factors, showing that SR protocol is the fastest among all protocols that are suitable for homogeneously-broadened transitions (Fig.~\ref{fig:performance}b). Furthermore, Fig.~\ref{fig:performance}c shows scaling of the optimal peak control intensity ($\propto \Omega_{\rm C}^{2}$) as a function of probe bandwidth $B$, with $\Omega_{\rm ATS}$ = $2\pi (2B)$; $\Omega_{\rm EIT}$ = $2\pi (6.6B)$; and $\Omega_{\rm SR}$ = $2\pi (31.2B)$,
implying orders of magnitude larger control-power is required for the SR protocol. In sum, SR memory operates optimally at substantially lower optical depths, but its demand on control-power is significantly larger.
%%%%%%%%%%%%%%%%%% 2-3 sentences on the impact of four-wave-mixing noise on the storage fidelity %%%%%%%%%%%%%%%%%%%%%%%%%%%%%%%%%%%%%%
\textcolor{black}{A protocol's resource dependence (on $d$ and $\Omega_{\rm C}$) plays a crucial role in determining the impact of fundamental noise processes on the storage fidelity. In particular, photonic four-wave-mixing (FWM) noise bears an exponential and quadratic scaling to the optical depth and control power, respectively~\cite{Lauk2013,Romanov2016,Geng2014}, and as such, the FWM strength associated with an optimal SR memory can be a factor of 3-7 times lower than that for an adiabatic memory~\cite{saglamyurek2021storing,supplementary}.}

In conclusion, we experimentally demonstrated a broadband spin-wave memory based on the superradiance effect in a cold Rb gas. This memory approach offers the shortest pulse storage in systems with transition linewidths narrower than the signal bandwidth. Besides conventional platforms like atomic gases and solid state systems, a high-performance SR memory could be implemented using Bose-Einstein condensates in the originally conceived superradiant regime with large density and dimensions comparable to the excitation wavelength~\cite{PhysRev.93.99}, and other interesting effects may emerge by considering subwavelength structured arrays when applied to memory~\cite{Asenjo-Garcia2017,Ballantine2020}.  
Beyond broadband quantum memories, our results may find use in the realizations of fast and efficient heralded single photon sources~\cite{ho2018optimal}, atom-based optical processing~\cite{Mazelanik2019}, and superradiance-mediated dipole blockade effects~\cite{PhysRevLett.125.073601,PhysRevLett.125.073602}.

\acknowledgements{This work was supported by the University of Alberta; the Natural Sciences and Engineering Research Council, Canada (Grants No.\ RGPIN-04523-16, No.\ RGPIN-2014-06618, and No.\ CREATE-495446-17);  the Alberta Quantum Major Innovation Fund; Alberta Innovates; and the Canada Research Chairs (CRC) Program. As researchers at the University of Alberta, we acknowledge that we are located on Treaty 6 territory, and that we respect the histories, languages, and cultures of First Nations, M\'etis, Inuit, and all First Peoples of Canada, whose presence continues to enrich our vibrant community.}

%\bibliography{library_arxiv.bib}
%Control: production of eprint (-1) disabled
%

%%%% Supplementary Information
\pagebreak
\pagebreak
\newpage
\newpage
\begin{widetext}

\begin{center}
% \textbf{\large SUPPLEMENTARY INFORMATION}
{\large \bf Supplemental Material for ``Superradiance-mediated photonic storage for broadband quantum memory''}

\end{center}
\end{widetext}

%%%%%%%%%% Merge with supplemental materials %%%%%%%%%%
%%%%%%%%%% Prefix a "S" to all equations, figures, tables and reset the counter %%%%%%%%%%
\setcounter{equation}{0}
\setcounter{figure}{0}
\setcounter{table}{0}
\setcounter{page}{1}
\setcounter{section}{0}
\makeatletter
\renewcommand{\theequation}{S\arabic{equation}}
\renewcommand{\thefigure}{S\arabic{figure}}
\renewcommand{\thetable}{S\arabic{table}}
\renewcommand{\thesection}{S\Roman{section}}
%\renewcommand{\thesubsection}{s\arabic{subsection}}
%\renewcommand{\bibnumfmt}[1]{[S#1]}
%\renewcommand{\citenumfont}[1]{S#1}
%%%%%%%%%% Prefix a "S" to all equations, figures, tables and reset the counter %%%%%%%%%%

\section{Evolution of coherences in superradiance protocol} \label{SI:sec1}
The operation of superradiance protocol (Fig.~1a-d) is based on resonant interaction between an optically dense atomic ensemble and two electromagnetic fields, the ``probe'' and ``control''. In the main text, we discuss its implementation in a $\Lambda$-type atomic system in terms of the exchange of coherences between the photonic ($|E(z,t)|^{2}$) and the atomic \big($|P(z,t)|^{2}$, $|S(z,t)|^{2}$\big) modes, using the Maxwell-Bloch equations
\begin{align}
\partial_z \hat{E}(z,t) & =  i\frac{g \sqrt{N}}{c} \hat{P}(z,t),\label{eq:MB1}\\
\partial_t \hat{P}(z,t) &= - \gamma \hat{P}(z,t) \!+\! i g \sqrt{N} \hat{E}(z,t) + \frac{i}{2} \Omega_{\rm c}(t) \hat{S}(z,t),\label{eq:MB2}\\
\partial_t \hat{S}(z,t) &= \frac{i}{2} \Omega_{\rm c}^*(t) \hat{ P}(z,t), \label{eq:MB3}
\end{align}
where we consider the 1D-propagation of optical fields along the medium from $z = \{0,L\}$. The photonic coherence $\hat{E}(z,t)$ describes the spatial and temporal variation of the probe field, while $\hat{P}(z,t)$ is the macroscopic polarization coherence of the $\ket{g} \leftrightarrow \ket{e}$ transition  and $\hat{S}(z,t)$ is the spin-wave atomic coherence of the $\ket{e} \leftrightarrow \ket{s}$ transition. The atom-probe coupling strength, in terms of optical depth, is $g\sqrt{N}=\sqrt{{c d \gamma}/{2L}}$, and we assume negligible relaxation between the levels $\ket{g}$ and $\ket{s}$.

Following the mathematical framework of~\cite{vivoli2013high,ho2018optimal}, we obtain analytic expressions for $P(z,t)$ and $S(z,t)$ during the absorption, writing, and retrieval stages of the protocol. Further, since Maxwell-Bloch equations are applicable to both weak classical fields (inducing macroscopic atomic polarization) and single-photon fields (inducing a single excitation in the ensemble), in the following analyses, we drop the operator notation from the corresponding variables, treating them as fields and macroscopic observables.

\subsection{Absorption stage: generation of atomic polarization} \label{sec:absorb}
To understand how the atomic coherence is initially established, we examine the temporal and spatial profiles of the macroscopic polarization $P(z,t)$ as it is generated in the medium during the absorption of input probe. In the absence of a control field ($\Omega_{\rm C} = 0$), equations (\ref{eq:MB1})-(\ref{eq:MB3}) reduce to
\begin{align}
&\partial_z E(z,t) =  \frac{i g \sqrt{N}}{c} P(z,t),\label{eq:SMB1}\\
&\partial_t P(z,t) = - \gamma P(z,t) \!+\! i g \sqrt{N} E(z,t), \label{eq:SMB2} 
\end{align}
with initial conditions given by the input photonic mode's shape: $E(z=0,t) = \mathcal{E}_{\rm in}(t) = E_{\rm P} e^{t/T} u(-t)$. Here, $\mathcal{E}_{\rm in}(t)$ is the electric field of the input probe, which
has an amplitude $\mathcal{E}_{\rm P}$ and a $1/e$ duration of $T$ and $u(-t)$ is the Heaviside 
unit step function. As discussed in the main text, this exponentially-rising temporal shape optimises both the amplitude as well as the spatial distribution of the polarization build-up, since it corresponds to the time-reversed impulse-response of the cold atomic ensemble which is characterized by a Lorentzian lineshape in this study. 

With the above definition of the input pulse and setting $t = 0$ as the time at which the input is suddenly shut off, we consider the scenario in which absorption and the associated polarization build-up occurs for times $t \le 0^{-}$, and for which subsequent re-emission or storage can occur at times $t \ge 0^{+}$. We assume that the probe field maintains its temporal shape during propagation through the medium ($c T \gg L$) and that it is only changed by an attenuation of its amplitude. We also note that in the main text, $T_{\rm P}$ refers to the $1/e$ duration of probe \emph{intensity}, such that $T_{\rm P} = T/2$. 

Next, to study the temporal dynamics of the polarization, we convert Eqs.~(\ref{eq:SMB1})-(\ref{eq:SMB2}) from the spatial to the complex spatial-frequency domain by taking the Laplace transform $z \rightarrow u$ and denoting the transformed functions with ``bars'': $\bar{X}(u,t) = \mathcal{L}\left[X(u,t)\right] = \int dz e^{-uz} X(z,t)$. In the transformed basis, the equivalents of (\ref{eq:SMB1}) and (\ref{eq:SMB2}) are 
\begin{align}
 \bar{E}(u,t) &=  \frac{i g \sqrt{N}}{c u} \bar{P}(u,t) + \frac{1}{u}\mathcal{E}_{\rm in}(t) \label{eq:SMB3} \\
\partial_t \bar{P}(u,t) &= - \gamma \bar{P}(u,t) + i g \sqrt{N} \bar{E}(u,t). \label{eq:SMB4} 
\end{align}
Inserting (\ref{eq:SMB3}) into (\ref{eq:SMB4}), we obtain a first-order linear differential equation in $\bar{P}(u,t)$
\begin{align}
& \dot{\bar{P}}(u,t) + A \bar{P}(u,t) =   \frac{i g \sqrt{N}}{u} \mathcal{E}_{\rm in}(t), \label{eq:SMB5}
\end{align}
where 
\begin{align}
A \equiv \gamma + \frac{g^{2} N}{c u} = \gamma\left(1 + \frac{d}{2 L u}\right)  
\end{align}
 and $d = {g^2 N L}/{2 \gamma c}$ is the on-resonant optical depth. Solving (\ref{eq:SMB5}) analytically, we find 
\begin{align}
& \bar{P}(u,t) = \bar{P}(u,0)  e^{-A t} + \frac{i g \sqrt{N} T}{u (1 + AT)} \mathcal{E}_{\rm in}(t). \nonumber
\end{align}
Since polarization is induced only by the probe field, and no atoms are excited initially, $\bar{P}(u,0) = 0$ and
\begin{align}
& \bar{P}(u,t) = \frac{i g \sqrt{N} T}{u (1 + AT)} \mathcal{E}_{\rm in}(t). \label{eq:SMB6}
\end{align}
This expression describes generation of the polarization that is in phase with the input photonic field for a given spatial position $z \rightarrow u$ (eg: see Fig.~1c for the polarization profile at z = 0). While deriving (\ref{eq:SMB6}), we have assumed that the probe switch-off at time $t = 0$ is instantaneous (i.e., the finite fall-time is neglected) such that the peak value of $\bar{P}$ occurs at $\bar{P}(u,t = 0)$, when the probe amplitude is maximized.

Next, we consider the spatial variation of the polarization as the probe propagates through the atomic medium: we expect that the polarization, because it is  in-phase with probe field, will undergo attenuation during propagation. Using an approach very similar to the one above, we model the spatial variation of $P$ along the propagation direction $z$ at a given instant of time. Taking the Laplace transform in the time domain $t \rightarrow \omega$ of equations~(\ref{eq:SMB1}) and (\ref{eq:SMB2}), with tildes denoting the transformed variables, we find
\begin{align}
&\partial_z \widetilde{E}(z,\omega) =  \frac{i g \sqrt{N}}{c} \widetilde{P}(z,\omega),\label{eq:SMB7}\\
&\widetilde{P}(z,\omega) = \frac{1}{\omega + \gamma}\left[ig \sqrt{N} \widetilde{E}(z,\omega) \:+ P(z, t = -\infty) \right]. \label{eq:SMB8} 
\end{align}
Since the medium does not contain any initial polarization, $P(z, -\infty) = 0$ and equation~(\ref{eq:SMB8}) can be substituted into (\ref{eq:SMB7}) to obtain:
\begin{align}
& \partial_z \widetilde{E}(z,\omega) = -\frac{g^2 N}{c (\omega + \gamma)} \widetilde{E}(z,\omega) \nonumber,
\end{align}
yielding
\begin{align}
& \widetilde{E}(z,\omega) = e ^{-{g^2 N}z/{c (\omega + \gamma)}} \widetilde{E}(z=0, \omega) \label{eq:SMB9}
\end{align}
Equation~(\ref{eq:SMB9}) represents the exponential decay in the probe-field amplitude along the propagation direction. Inserting it into (\ref{eq:SMB8}) gives the spatially decaying polarization profile
\begin{align}
& \widetilde{P}(z,\omega) = \frac{i g \sqrt{N}}{\omega + \gamma} e^{-{g^2 Nz}/{c(\omega + \gamma)}} \widetilde{E}(z=0, \omega) \label{eq:SMB10}
\end{align}
Thus, at a fixed position $z$ within the medium, the polarization amplitude grows in time, following the envelope of the probe field (\ref{eq:SMB6}), whereas at any fixed instant in time, the polarization amplitude exhibits an exponential decay along the medium length (\ref{eq:SMB10}).

\subsection{Writing stage: Polarization-to-spin-wave conversion} \label{sec:storage}
During absorption, the input-probe field imprints its phase pattern onto the atomic ensemble in the form of a polarization build-up over the entire probe duration. By applying a "write" control field immediately after the probe is absorbed (and when the photonic coherence has been completely converted to the polarization coherence), this atomic polarization is transferred into a long-lived spin-wave coherence of the atomic ensemble (Fig.~1a,c). In our formalism, absorption is maximized at $t = 0$ with the value
\begin{align}
& \bar{P}(u,0) = \frac{i g \sqrt{N} T}{u (1 + AT)} \mathcal{E}_{\rm p} \label{eq:SMB11},
\end{align}
which serves as the  source term upon which the write control acts. The transformation between the polarization and spin coherences is governed by the Maxwell-Bloch equations (\ref{eq:MB1})-(\ref{eq:MB3}), written again here for convenience:
\begin{align}
& \partial_z E(z,t) =  \frac{i g \sqrt{N}}{c} P(z,t),\label{eq:SMB12} \\
&\partial_t P(z,t) = - \gamma P(z,t) \!+\! i g \sqrt{N} E(z,t) + \frac{i}{2} \Omega_{\rm c}(t) S(z,t),\label{eq:SMB13} \\
&\partial_t S(z,t) = \frac{i}{2} \Omega_{\rm c}(t) P(z,t), \label{eq:SMB14}
\end{align}
where we have assumed $\Omega_{\rm C}$ to be real and ignored any spatial depletion of the control intensity (i.e.~$\Omega_{\rm C} (z , t) = \Omega_{\rm C} (0 , t)$).~We consider a square control pulse 
\begin{align}
& \Omega_{\rm c}(t) = \Omega_{\rm c}, \:\:\:\: 0 \le t \le T_{\rm c}. \label{eq:SMB15} 
\end{align}

As the $P \rightarrow S$ mapping occurs over a time-interval given by the control duration, $T_{\rm C}$ must be much shorter than the superradiant emission time $T_{\rm SR}$, or else a significant portion of the built-up polarization is lost via superradiant emission. Taking the Laplace transform of (\ref{eq:SMB12})-(\ref{eq:SMB14}) from $z \rightarrow u$:
\begin{align}
&\bar{E}(u,t) = \frac{i g \sqrt{N}}{cu} \bar{P}(u,t) + \frac{1}{u}E(z = 0 , t) \label{eq:SMB16} \\
&\partial_t \bar{P}(u,t) = - \gamma \bar{P}(u,t) \: + i g \sqrt{N} \bar{E}(u,t) \: + \frac{i}{2} \Omega_{\rm c}(t) S(u,t) \label{eq:SMB17} \\
&\partial_t \bar{S}(u,t) = \frac{i}{2} \Omega_{\rm c} \bar{P}(u,t) \label{eq:SMB18},
\end{align}
Since no photonic field $E$ is present in the medium at the time when the write control is turned on ($t = 0^{+}$), $E(z = 0 , t) = 0$. Next, substituting (\ref{eq:SMB16}) and (\ref{eq:SMB18}) into the time-derivative of (\ref{eq:SMB17}), we obtain a second-order linear differential equation in $\bar{P}(u,t)$
\begin{align}
& \ddot{\bar{P}}(u,t) + A\dot{\bar{P}}(u,t) + B\bar{P}(u,t) = 0, \label{eq:SMB19}
\end{align}
where $A \equiv \gamma + g^{2} N/c u = \gamma(1 + {d}/{2 L u})$ and $B = \Omega_{\rm c}^{2}/4$. The roots of the characteristic equation in (\ref{eq:SMB19}), can be real or complex depending on the sign of the discriminant $D = A^{2} - 4B$ which is determined by the relative strengths of $A$ and $B$ coefficients. For the superradiance protocol the condition for fast storage and retrieval requires a strong driving control such that
\begin{equation}
\Omega_{\rm C} \gg \gamma(1 + d), \nonumber
\end{equation}
implying $\Omega_{\rm c} \gg \gamma(1 + {d}/{2 L u})$ and thus, $4B \gg A^2$. Roots of the characteristic equation are thus complex \big($-{A}/{2} \pm i{\Omega_{\rm C}}/{2}$\big) giving a solution of the form
\begin{align}
& \bar{P}(u,t) = e^{-{At}/{2}}\left[\:C(u) \cos\left(\frac{\Omega_{\rm c}t}{2}\right) + D(u) \sin\left(\frac{\Omega_{\rm c}t}{2}\right)\:\right] \nonumber
\end{align}
for $t \ge 0$.  

Applying the initial condition at $t = 0$ (\ref{eq:SMB11}), the coefficient $C(u) = P(u,t=0)$. Also, the fact that $P(u,t = T_{\rm C})$ must ideally be 0 by the end of the control pulse suggests $D(u)=0$. The expression for $\bar{P}(u,t)$ during the writing stage is thus,
\begin{align}
\bar{P}(u,  0 \le t \le T_{\rm c}) = e^{-{At}/{2}}\cos\left(\frac{\Omega_{\rm c}t}{2}\right) P(u,t = 0). \label{eq:SMB20}
\end{align}
As can be seen, for $\Omega_{\rm C} T_{\rm C} = \pi$ (i.e., a $\pi$-control-pulse with duration $T_{\rm C}$), no polarization remains in the medium by the end of control duration (i.e., $\bar{P}(u,T_{\rm C}) = 0$) and is completely converted into the spin-wave $\bar{S}(u,t = T_{\rm C}) = \frac{i \Omega_{\rm C}}{2}\int_{0}^{T_{\rm C}} \bar{P}(u,t')\: dt'$:
\begin{align}
\bar{S}(u,T_{\rm C}) = \frac{i \Omega_{\rm C}}{A^2 + \Omega_{\rm C}^2}\bar{P}(u,0)\big[\Omega_{\rm C} \:e^{-A T_{\rm C}/2} + A\big] \label{eq:SMB21}
\end{align}
Now, since we are in the regime of $\Omega_{\rm C} \gg A$: $\pi/T_{\rm C} \gg A$ or $A T_{\rm C}/2 \ll 1$, giving $e^{-A T_{\rm C}/2} \approx 1$. Equation~(\ref{eq:SMB21}) thus simplifies to:
\begin{align}
\bar{S}(u,T_{\rm C}) = iP(u,t=0), \label{eq:SMB22}
\end{align}
indicating that a lossless conversion of the polarization coherence into the spin-wave coherence can be achieved in the fast storage regime using a $\pi$-control-pulse.  

\subsection{Converting the stored spin-wave back into polarization during retrieval} \label{sec:retrieval}
In previous subsection, we saw that a fast $\pi$-pulse can quickly act on the atomic polarization to convert it into the spin-wave with a unity-transfer-efficiency, thus effective storage of the input photonic coherence. The probe field remains stored as the spin-wave up to a time, referred to as the memory lifetime ($T_{\rm Mem}$), governed by the spin-wave decoherence processes. Re-applying the $\pi$ pulse after a desired storage interval $(T_{\rm S} \le T_{\rm Mem})$, initiates the reversed mapping of $S$ into $P$ which then couples to the output photonic field (Fig.~1b,d).~Assuming the absence of any decoherence mechanism that would otherwise degrade the spin-wave amplitude, we have the initial condition as $\bar{S}(u, \tau) = \bar{S}(u,T_{\rm C})$ where $\bar{S}(u,T_{\rm C})$ is the spin-wave generated by the end of the writing stage and $\tau = T_{\rm C} + T_{\rm S}$ with all times measured relative to the the end of absorption process ($t = 0$).~The control field for readout is defined as
\begin{align}
& \Omega_{\rm c}(t) = \Omega_{\rm c}, \:\:\:\: \tau \le t\le \tau + T_{\rm c}, \label{eq:SMB23} 
\end{align}
so that the spin-to-polarization transfer occurs within the short duration of $T_{\rm C}$.~Taking the time derivative of (\ref{eq:SMB18}) and then inserting (\ref{eq:SMB16}) and (\ref{eq:SMB17}), we get a second order differential equation in $\bar{S}(u,t)$
\begin{align}
& \ddot{\bar{S}}(u,t) + A\dot{\bar{S}}(u,t) + B\bar{S}(u,t) = 0, \label{eq:SMB24}
\end{align}
where we used the fact that no photonic field is initially present at the time of readout; the $A$ and $B$ coefficients are defined as before.~The condition for fast retrieval yields $4B \gg A^{2}$, giving a solution of the form
\begin{align}
&\bar{S}\big(u, \tau \le t \le \tau + T_{\rm C}\big) = e^{-A t/2}\cos\bigg(\frac{\Omega_{\rm c}t}{2}\bigg) \bar{S}\big(u,\:t = T_{\rm C}\big). \label{eq:SMB25}
\end{align}
Again for a control pulse such that $\Omega_{\rm C} T_{\rm C} = \pi$ during the interval $t = T_{\rm C}$, the spin-wave in (\ref{eq:SMB25}) goes to 0 and is completely converted into the polarization coherence using $\bar{P}(u,\: t) = \big(2/i \Omega_{\rm c}\big) \partial_t \bar{S}(u,t)$:
\begin{align}
& \bar{P}(u,t) = \frac{i e^{-At/2}}{\Omega_{\rm c}}\bigg[\Omega_{\rm c} \sin \bigg(\frac{\Omega_{\rm c} t}{2} \bigg) + A \cos \bigg(\frac{\Omega_{\rm c} t}{2}\bigg) \bigg] \nonumber \\
&\times \bar{S}\big(u , \: t = T_{\rm C}\big),\nonumber
\end{align}
which, for a $\pi$-pulse (and $e^{-A T_{\rm C}/2} \approx 1$) simplifies to
\begin{align}
\bar{P}(u,\: \tau + T_{\rm c}) = i\bar{S}(u,\: t = T_{\rm C}),
\end{align}
indicating a lossless conversion of spin-wave into polarization at the time of readout. 

\section{Spectral bandwidth of input probe} \label{sec:fourier}
\textcolor{black}{We derive the Fourier-limited relationship between the rise time $T_{\rm P}$ and the bandwidth at full-width-half-maximum $B$ of an excitation probe that has an exponentially rising temporal intensity profile of the form:
\begin{align}
& I(t) = e^{t/T_{\rm P}} u(-t) + e^{-t/T_{\rm F}} u(t) , \label{eq:probe}
\end{align}
where $T_{\rm P}$ and $T_{\rm F}$ are, respectively the $1/e$ rise and fall times of the probe intensity, such that ideally $T_{\rm F} \ll T_{\rm P}$.~The FWHM duration of such a pulse is given by $T_{\rm FWHM} = (T_{\rm P} + T_{\rm F}) \ln(2)$.~We determine the spectral width of the probe in terms of the 3 dB frequency range ($B = \omega_{3 dB} / 2\pi$) using the Fourier transform of the probe electric field $E(t) = e^{t/2T_{\rm P}} u(-t) + e^{-t/2T_{\rm F}} u(t)$, given by
\begin{align}
& E(\omega) = \frac{1}{2\pi} \Bigg[\int_{-\infty}^{0} e^{\big(\frac{1}{2T_{\rm P}} - i\omega \big)t} + \int_{0}^{+\infty} e^{- \big(\frac{1}{2T_{\rm F}} + i\omega \big)t} \: \Bigg] dt \nonumber \\
&           = \frac{T_{\rm P} + T_{\rm F}}{\pi}\bigg[\frac{1}{(1 + 4T_{\rm p}T_{\rm F}\omega^{2}) - 2i(T_{\rm P} - T_{\rm F})\omega} \bigg]
\label{eq:spectral probe}
\end{align}
The intensity spectral profile, computed using $I(\omega) = E^{*}(\omega)E(\omega)$ is 
\begin{align}
& I(\omega) = \bigg(\frac{T_{\rm P} + T_{\rm F}}{\pi}\bigg)^{2}\bigg[\frac{1}{(1 + 4T_{\rm P}T_{\rm F}\omega^2)^{2} + 4(T_{\rm P} - T_{\rm F})^2\omega^2}\bigg], \label{eq:probe intensity spectral}
\end{align}
which can then be numerically plotted to estimate $\omega_{3 dB}$. As an example, for numerical simulations in Figs. 1 and 3 of the main text, we chose $T_{\rm F} = \frac{1}{10} T_{\rm P}$. Substituting in (\ref{eq:probe intensity spectral}), we obtain $\omega_{3 dB}T_{\rm FWHM}$ = 0.75 or $B = 0.157/T_{\rm P}$ with $T_{\rm FWHM} = 0.76T_{\rm P}$. In our experiments, the fall time $T_{\rm F}^{\rm exp}$ is limited by the electronic response of the arbitrary waveform generator driving the probe-shaping EOM, and was measured to be $\approx$ 6 ns. Therefore, probe pulses with $T_{\rm P} =$ 20 ns and $T_{\rm F}^{\rm exp} = \frac{1}{3}T_{\rm P}$ (Fig.~2 b,c in main text) correspond to $B = 0.147/T_{\rm P}$ = 7.4 MHz which is equivalent to 1.23$\Gamma/2\pi$ for Rb atoms. Similarly, probe pulses with $T_{\rm P} =$ 10 ns (Fig.~2d,e,f) correspond to $B = 0.13/T_{\rm P} =$ 12.7 MHz $= 2.12\Gamma/2\pi$.}

\section{Experimental Setup and measurements} \label{sec:experiment}
\subsection{Cold Atom Preparation} \label{sec:ensemble}
An ensemble of laser-cooled $^{87}$Rb atoms serves as the medium for either probing the superradiant emission (Fig.~2b,c) or for storing short optical pulses via the superradiance (SR) protocol (Fig.~2d,e,f).~The atoms are \textcolor{black}{loaded from a background atomic vapour} into a 3-dimensional retro-reflected magneto-optical-trap (MOT) operating on the D2 line.~The MOT repump laser is frequency locked to the $\ket{F=1} \rightarrow \ket{F'=2}$ transition using saturated absorption spectroscopy, at a detuning of -80 MHz. The repump beam is upshifted by +80 MHz using an acousto-optic-modulator (AOM) to make it resonant with the repump transition.~The cooling (trapping) laser is frequency stabilized with respect to the repump laser using the offset beatnote locking technique such that it is detuned by $\approx$ -180 MHz from the $\ket{F=2} \rightarrow \ket{F'=3}$ transition.~The cooling beam is passed through a double-pass AOM stage which upshifts the frequency by +160 MHz, making it red-detuned from the cooling transition effectively by $\approx$ 3.5$\Gamma$.~The 4 seconds of MOT cooling is followed by 5 ms of polarization-gradient optical molasses, at the end of which about $N_A = (2.0 \pm 0.5)\times 10^{\rm 8}$ Rb atoms are prepared in $\ket{F=2}$.~\textcolor{black}{Time-of-flight (TOF) \textcolor{black}{absorption-imaging} measurements yield an elliptical-shaped cloud with $1/e^2$ Gaussian diameters of 2.5 mm and 3.6 mm along the radial direction, and a cloud temperature of $\approx$ 35 $\mu$K}. A 3.5 ms optical-pumping stage follows molasses, wherein the repump beam is turned off but the cooling beam is kept on to off-resonantly transfer atoms from $\ket{F=2}$ into the $\ket{F=1}$ state, which serves as the initial population level \big($\ket{g}$\big) for the input probe field. \textcolor{black}{Absorption-and-reemission measurements of short, resonant, and weak optical probes yield an ensemble optical depth of $d =$ 9 $\pm$ 1}.~For datasets like the one in Fig.~2c, a variable time-of-flight spatial expansion stage was introduced after the optical pumping stage, to controllably vary this optical depth. 

\subsection{Generation of probe and control fields} \label{sec:beams}
In our experiments, the probe and the control fields resonant with $\ket{F=1} \rightarrow \ket{F'=2}$ and $\ket{F=2} \rightarrow \ket{F'=2}$ transitions are derived from the MOT repump and cooling lasers respectively.~To generate the probe pulse at the desired frequency, the CW beam from the repump laser is both gated and frequency-shifted by -80 MHz using an AOM, such that the gated pulse is -160 MHz away from the probe transition. This gated pulse (200 ns wide) is subsequently passed through a double-pass AOM which imparts a frequency shift of +160 MHz, thereby producing a resonant probe pulse.~This resonant probe is then passed through an electro-optic intensity modulator (EOM) which shapes the gated pulse into an exponentially-rising temporal envelope with the desired rise time ($T_{\rm P}$).~The CW cooling-laser-beam is passed through an AOM that generates Gaussian write and read temporal control pulses ($T_{\rm C} =$ 20 ns at FWHM), and imparts an additional frequency shift of -85 MHz making them resonant with the control transition.~The RF oscillators driving the gated-probe and control AOMs are amplitude controlled using one two-channel arbitrary waveform generator (AWG), while the probe-intensity EOM is driven directly from a separate AWG.~The center frequency of probe pulse is set by adjusting the carrier frequency of the probe-double-pass AOM, while that of control pulse is set by changing the beatnote offset frequency between the repump and cooling lasers. AWGs are also used to adjust the relative timing positions between the probe and control fields.

After setting the amplitude (power), frequency, and timing of probe and control pulses, both beams are launched separately into the MOT chamber \textcolor{black}{with the control field oriented at an angle ($\theta$) of either 5$^{\circ}$ or 50$^{\circ}$ with respect to the probe-axis}.~The probe is focused at the center of the atomic cloud with a $1/e^{2}$ Gaussian diameter of 70 $\mu$m, while the control is collimated at 2.5 mm diameter.~The initially linear polarization of each field is adjusted using $\lambda/4$-waveplates to maximise the atom-light interaction strength.

\subsection{Measurement and detection} \label{measure}
The useful signals at the output of the probe arm include the input probe photons (in the absence of both atoms and control), superradiantly emitted probe photons (with atoms but without control), and stored-and-retrieved photons (with both atoms and control).~We detect these signals using time-resolved photon counting measurements, enabled by a single-photon-detector (SPD) and a time-to-digital converter (TDC), in the form of histograms of the detected photon counts with respect to their arrival times.~The SPD-TDC combination provided both a better temporal resolution as well as a lower electronic noise-floor for our detection signals compared to a photodiode + oscilloscope pair.

In a single experimental cycle, after preparing the cold ensemble, we perform measurement-and-detection during a 1 ms window by sending a sequence of 1000 probe pulses (also 1000 write-read control pulses when implementing memory) with a 1 $\mu$s-separation between consecutive pulses. The measurement-and-detection stage is followed by a 200 ms of `wait-time' before the next experimental cycle begins. By repeating several such prepare-and-measure cycles ($n_{\rm cyc}$), signal integration over a large number of trials is obtained ($n_{\rm cyc}\times 10^3$). 

\textcolor{black}{The probe pulses injected into the atomic cloud either contain large input mean photon numbers per pulse (i.e., $\bar{n}_{\rm in}\sim 10^3$) or are at the single-photon-level where $\bar{n}_{\rm in}$ can be much less than one. In both cases, we estimate $\bar{n}_{\rm in}$ by measuring the number of probe photons detected ($N_{\rm D}$) with respect to the number of probe pulses generated over the same time-interval ($N_{\rm P} = n_{\rm cyc}\times 10^3$) and in the absence of atoms, such that $\bar{n}_{\rm in}= \big(1/\eta_{\rm T}\big)\times \big(N_{\rm D} / N_{\rm P}\big)$, where $\eta_{\rm T} \approx 0.25$ is the overall detection efficiency including the end-to-end transmission efficiency of the optical setup, fiber-coupling efficiency, and SPD efficiency.~For measurements with $\bar{n}_{\rm in} \gg 1$, a series of neutral-density-filters providing up to 38 dB optical-attenuation are placed along the output-path of the probe arm. This ensures that in spite of large $\bar{n}_{\rm in}$, the effective number of photons reaching the SPD remain below 1 so that the detector does not go into the saturation mode}

\textcolor{black}{Superradiance memory using $\bar{n}_{\rm in} \gg 1$ probe is implemented at the $\theta = 5^{\circ}$ setting whereas the single-photon-level memory measurements ($\bar{n}_{\rm in} < 1$) are taken at the $\theta = 50^{\circ}$ setting. Limited by the power in our setup, the shortest write-and-read control pulses we could generate with a pulse area close to $\pi$ correspond to a Gaussian temporal shape with an FWHM duration of $T_{\rm C} =$ 20 ns and a peak power of 8 mW.}

\subsection{Probe absorption-linewidth and peak optical-depth} \label{sec:Linewidth and OD}
\begin{figure}
\begin{center}
\includegraphics[width = 85 mm] {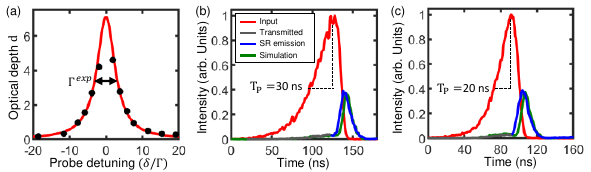}
\caption{\textbf{(a)} Absorption linewidth of the probe transition ($\Gamma^{\rm exp}$).~Black circles are the measured optical depth values with solid line as fit to the data (\ref{eq:OD fit}). \textbf{(b,c)} Estimation of peak optical depth by simulating the absorption-and-reemission measurements of short exponential probes using $\Gamma^{\rm exp}$, yielding the peak optical depth of 9.} 
\label{fig:OD}
\end{center}
\end{figure}

To characterize the $\ket{5S_{1/2}, \: F = 1} \leftrightarrow \ket{5P_{3/2}, \: F' = 2}$ probe transition we perform spectral measurements of the absorption linewidth ($\Gamma^{\rm exp}$) and the resonant optical depth ($d_{\rm res}$) for our Rb ensemble.~Comparison of $\Gamma^{\rm exp}$ with respect to the natural linewidth would indicate whether there are additional broadening mechanism (eg: Doppler broadening) in our system.~We send a sequence of 1000 square probe pulses, of duration 400 ns, at a fixed detuning ($\delta$) and detect the probe transmission counts both without \big[$I_{\rm in}(\delta)$\big] and with \big[$I_{\rm out}(\delta)$\big] atoms.~By performing such sequence over many cycles, reliable statistics can be obtained. The relatively long duration of the probe ($T_{\rm P} = 3.7/\gamma$) ensures no reemission occurs after absorption. These measurements are performed for values of $\delta$ between -19 MHz to + 19 MHz with the optical depth in each case computed as $d \: (\delta) = -ln \: (I_{\rm out} / I_{\rm in})$. By fitting a Lorentzian to the measured $d$ values
\begin{align}
 d (\delta) = \frac{d_{\rm res}}{1 + 4\left(\delta/\Gamma\right)^{2}} ,  \label{eq:OD fit}
\end{align}
we extract the absorption linewidth $\Gamma^{\rm exp} = 2\pi \times$\big(6.2 $\pm$ 0.1 MHz\big), and the peak optical depth $d_{\rm res} =$ 7 $\pm$ 0.2 (Fig.~\ref{fig:OD}a). We note that the maximum measured OD from our data is limited to $\approx$ 5.6 and hence less than what the fit in (\ref{eq:OD fit}) gives.~We attribute this to the limited dynamic range of the SPD and thus exclude the data points near $\delta = 0$ while performing the fit.~The estimated value of $\Gamma^{\rm exp}$ agrees very well with the natural linewidth, confirming negligible effect of any inhomogeneous broadening and that the condition $2\pi B > \Gamma^{\rm exp}$ is satisfied for observing superradiant emission, as in Fig.~2b.

Next, using the value of $\Gamma^{\rm exp}$, we simulate our measurements for the absorption-and-reemission of short probe pulses, shown in Fig.~\ref{fig:OD}b,c for $T_{\rm P} =$ 30 ns \big($0.55/\gamma$\big) and 20 ns \big($0.37/\gamma$\big).~A good agreement between experiments and simulations is obtained for a peak optical depth $d_{\rm res}=$ 9 $\pm$ 1 which is 25$\%$ higher than that obtained from spectral measurements.~Since, the spectral measurements are carried out with pulses that are an order-of-magnitude larger, depletion of the atomic cloud during each pulse sequence can account for the observed discrepancy.~Hence, the peak optical depth of our ensemble is closer to what we expect from the time-domain absorption measurements.

\subsection{Operational SR regime in our experiment} \label{sec:SR Regime}
\begin{figure}
\begin{center}
\includegraphics[width = 85 mm] {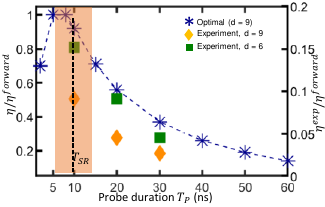}
\caption{Storage-and-forward-recall efficiency simulation for varying duration (bandwidth) of an exponentially rising probe using experimentally measured values of (d, $\gamma$, $T_{\rm SR}$) and optimized control. Probe durations close to $T_{\rm SR}$ yield optimal efficiencies (shaded area). Also shown are the experimental efficiencies under non-optimal control, for two different optical depths. The data follows the predicted trend that the SR protocol is favorable for storage of shorter pulses.} 
\label{fig:SR Regime}
\end{center}
\end{figure}
The SR protocol benefits from the generation of a collective polarization coherence at timescales shorter than the $T_{\rm 2} = 1/\gamma$ relaxation rate of the system and is inherently suited for storing pulses with duration $T_{\rm P} < 1/\gamma$. In this superradiant regime, an optimized probe duration is close to the superradiant emission time of the system i.e., $T_{\rm P}^{\rm opt} \approx T_{\rm SR}$, where $T_{\rm SR}$ is determined from the system parameters \big($d,\gamma$\big). Figure~\ref{fig:SR Regime} shows numerically calculated efficiencies $(\eta)$ relative to the optimal efficiency $(\eta^{\rm forward})$ for storage followed by forward-recall of an exponentially rising input probe of varying duration (bandwidth) and an abrupt switch-off ($T_{\rm F} = 0.1T_{\rm P}$) under our experimental configuration of $d =$ 9, $\gamma =$ 2$\pi$.~3 MHz, that yield $T_{\rm SR} =$ 8 ns.~The simulations consider square-shaped $\pi$-control pulses for writing and readout with duration $T_{\rm C}= \frac{1}{10}T_{\rm SR}$.~The shaded region depicts that for a given optical-depth, only a small range of probe durations around $T_{\rm SR}$ will yield optimal efficiency. The reduction in efficiency at lower values of $T_{\rm P}$ is due to inadequate optical-depth resulting in reduced probe absorption whereas at higher values, the drop is due to dephasing in the atomic dipoles \big($\propto e^{-\gamma T_{\rm P}}$\big) during absorption. 

Also shown are the experimentally measured efficiencies $(\eta^{\rm exp})$ for probe pulses with 1/e rise times of 10 ns, 20 ns, and 30 ns, each having a finite fall time of $\approx$ 6 ns, using our non-optimal and Gaussian-shaped control with $T_{\rm C} =$ 20 ns. Although, the $\eta^{\rm exp}$ values are much smaller than the expected $\eta^{\rm forward}$ due to the reasons discussed in the main text, they do follow the trend that the SR protocol readily favours the storage of shorter pulses and also confirm that our chosen parameters for the probe pulse to be stored lie in the good operational regime of the SR protocol at the optical depth that we have. 

\subsection{Optimizing the SR memory in our experiment} \label{sec:10 ns storage}
\begin{figure}
\begin{center}
\includegraphics[width = 85 mm] {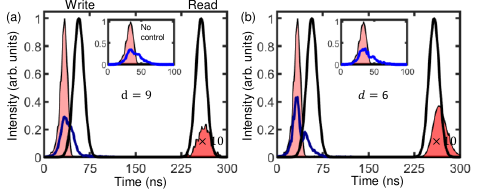}
\caption{Storage followed by on-demand retrieval of a probe with $T_{\rm P} =$ 10 ns using the SR protocol, at optical depths of \textbf{(a)} $d =$ 9 \big($T_{\rm SR} =$ 8 ns\big), and \textbf{(b)} $d =$ 6 \big($T_{\rm SR} =$ 11 ns\big). The measured overall memory efficiency values are 3 $\%$ and 5$\%$ respectively.} 
\label{fig:SR memory}
\end{center}
\end{figure}
Our experimental SR memory efficiency is primarily limited by the non-optimal writing stage: at $d = $ 9, the superradiant emission in our system drains the excited state within $T_{\rm SR}$ = 8 ns whereas the control-induced polarization-to-spin-wave transfer occurs over an interval six times longer, given by the full duration of our experimental write control ($2.25 \times T_{\rm C}$ = 45 ns). A comparison of probe's reemission profiles both in the absence and presence of the control field (Fig.~2d) suggests that a large fraction of the polarization coherence, accumulated during absorption, is lost via superradiant emission even when the control is present (referred to as ``leakage"). Faster writing requires shorter control duration and in turn more power to maintain the $\pi$-area. Limited by the control power, another way to alleviate this leakage is to extend the superradiant emission by lowering the system's optical depth.~Figure~\ref{fig:SR memory} illustrates the storage-and-retrieval of a 10 ns probe at optical depths of 9 and 6, where at the lower $d$, an increased superradiant emission time enables the control to transfer a greater fraction of the polarization coherence into spin-wave, thereby yielding a higher efficiency, even at the expense of increased probe transmission. Figure~\ref{fig:SR Regime} shows similar behaviour is observed for other probe durations. However, this optimization is limited by the trade-off between the transmission and leakage losses: lowering the $d$ further makes the probe transmission dominant and reduces the efficiency.

\subsection{Estimation of memory lifetime} \label{sec:memory lifetime}
Figure~2e in the main text shows the measured superradiance memory efficiency as a function of the storage time i.e., the time interval between the write and read control pulses. The dataset was obtained for the small-angle ($\theta = 5^\circ$) probe-control separation, wherein short exponential probe pulses of 10 ns duration were stored and then retrieved at later time intervals ranging between 200 ns to 6.2 $\mu$s. We recorded a memory efficiency of 3$\%$ for the storage duration of 200 ns. The efficiencies (black circles) at other storage times are normalized with respect to the above value. The degradation in the memory efficiency follows an exponential decay (solid red line) yielding with a 1/e memory lifetime of 4.2 $\pm$ 0.3 $\mu$s. This memory lifetime is limited by the dephasing of the collective spinwave excitation, induced by the residual / ambient magnetic fields. For the single-photon-level measurements done at the wider probe-control separation ($50^\circ$), the spinwave dephasing induced by the random atomic motion dominates since it is an increasing function of $\theta$ and limits the memory lifetime in our setup to 1.2 $\pm$ 0.2 $\mu$s.

%%%%%%%%%%%%%%%%%%%%%%%%%%%%5 Section on Single-photon-level measurements
\subsection{Single-photon-level measurements} \label{sec: single-photon-level}
To implement storage and recall of single-photon-level probe pulses ($\bar{n}_{\rm in} < 1$), we spatially filter out the noise from the scattered control photons leaking into the spatial mode of the memory retrieved signal. This is done by increasing the angle of the control beam with respect to the probe from $5^\circ$ to $50^\circ$ which increases the isolation from 80 dB to 120 dB (four-orders-of-magnitude extinction), whereas the memory efficiency reduces to 1.3$\%$ from the typical value of 3$\%$ for 200 ns storage time. 

We define $\rho_{\rm S}$ as the probability of detecting a `signal' count from the retrieved probe as $\rho_{\rm S} = \big(1/\eta_{\rm T}\big)\times \big(N_{S} / N_{\rm Mem}\big)$, where $N_{\rm S}$ is the total number of detection counts for the recalled probe after $N_{\rm Mem}$ memory trials, computed over a time window $\Delta t$ centered around the readout time. Analogously, we can define the probability of detecting a `noise' photon at the readout as $\rho_{\rm R} = \big(1/\eta_{\rm T}\big)\times \big(N_{R} / N_{\rm N}\big)$ where $N_{R}$ is the number of noise detections (within the signal readout window $\Delta t$) in the absence of the input probe (i.e., $\bar{n}_{\rm in} = 0$) after $N_{\rm N}$ trials. Both $\rho_{\rm S}$ and $\rho_{\rm R}$ are independently measured detection probabilities, with the memory efficiency ($\eta_{\rm Mem}$) and the ratio of the stored-and-recalled signal to noise (SNR) given as $\eta_{\rm Mem} = \rho_{\rm S} / \bar{n}_{\rm in}$ and $\text{SNR} = \rho_{\rm S} / \rho_{N}$.

\begin{figure}
\begin{center}
\includegraphics[width = 85 mm] {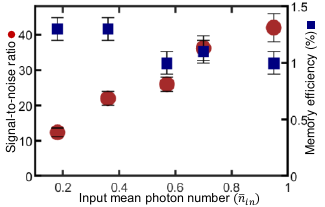}
\caption{Measured signal-to-noise ratio (SNR) and memory efficiency vs mean-photon
number ($\bar{n}_{\rm in} < 1$) for 10-ns-long probe pulses stored using SR memory for 200 ns.} 
\label{fig:SNR}
\end{center}
\end{figure}

Figure.~2f in the main article shows the storage and recall after 200 ns of a $T_{\rm P} = 10$ ns probe at $\bar{n}_{\rm in} = 0.18$, after $N_{\rm Mem} = 10^6$ attempts giving $N_{\rm S} =$ 645 $\pm$ 25 counts within a $\Delta t=$ 50 ns window yielding $\rho_{\rm S} =$ (2.6 $\pm$ 0.1) $\times 10^{-3}$ and $\eta_{\rm Mem} = (1.3 \pm 0.1) \%$. The inset of the readout window shows the noise contribution relative to the recalled probe (signal) under four configurations: (\rmnum{1}) ambient light + scattering from optics and atoms (both probe and control are off) ; (\rmnum{2}) leak from the input probe (control off); (\rmnum{3}) control without atoms and probe; and (\rmnum{4}) control and atoms without probe. Specifically, counts from (\rmnum{3}) and (\rmnum{4}) would indicate noise from photons that are added during the memory operation, such as from scattered control (\rmnum{3}) or any fundamental process (\rmnum{4}). We determine the unconditional noise probabilities of (6 $\pm$ 1)$\times 10^{-5}$, (7 $\pm$ 1)$\times 10^{-5}$, (1.3 $\pm$ 0.1)$\times 10^{-4}$, and (2.1 $\pm$ 0.2)$\times 10^{-4}$ respectively. The fact that both (\rmnum{3}) and (\rmnum{4}) configurations yielded similar $\rho_{\rm R}$ indicates that the scattered control remains the dominant noise source in our implementation and more importantly, there is no measurable contribution from any photonic noise linked to the operation of the superradiance memory. With the above noise contributions (all included in (\rmnum{4})), we obtain an SNR $=$ 12.4 $\pm$ 1.2. Figure~\ref{fig:SNR} shows SNR as a function of $\bar{n}_{\rm in}$ after 200-ns storage: while the memory efficiency stays roughly the same, the SNR shows a linear dependence.

%%%%%%%%%%%%%%%%%%%%%%%%%%%%%%%%%%%%%%%%%%%%%%%%%%%%%%%%%%%%%%%%%%%%%%%%%%

\begin{figure}
\begin{center}
\includegraphics[width = 85 mm] {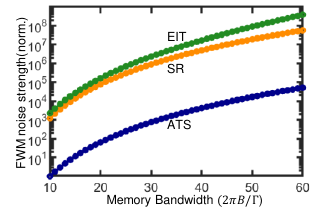}
\caption{Four-wave mixing (FWM) noise strength vs bandwidth, calculated for optimal ATS, SR, and EIT memories. The noise strength is normalized with respect to its minimum value corresponding to the smallest bandwidth of 10$\Gamma_{\rm eg}/2\pi$.} 
\label{fig:FWM}
\end{center}
\end{figure}

\section{Scaling of the four-wave-mixing photonic noise in SR, ATS, and EIT memories} \label{sec:FWM}
Photonic noise generated at the memory output via four-wave-mixing process is detrimental towards the reliable implementation of broadband spinwave storage-and-retrieval in $\Lambda$-type atomic media~\cite{Phillips2011,Michelberger2015,Lauk2013,Romanov2016}.~FWM noise is an important concern in memory configurations involving backward retrieval of the signal probe which on the one hand yield optimal memory efficiencies but on the other hand allow for perfect phase-matching among the optical fields involved in the FWM process~\cite{saglamyurek2021storing}.~In such a scenario, the impact of a phase-matched FWM noise corrupting the memory operation is characterized using a noise-strength parameter $(S_{\rm FWM})$ that depends on the optical-depth ($d$) and control Rabi frequency ($\Omega_{\rm C}$) as~\cite{Lauk2013,Geng2014,Romanov2016}:
\begin{equation}
    S_{\rm FWM}\propto \Omega_{\rm c}^4\left[\sinh\left(\frac{\zeta d \gamma_{\rm eg}}{2\Delta_{\rm gs}}\right)\right]^2, \label{eq:FWM}
\end{equation}
where $\big(\zeta, \gamma_{\rm eg}, \Delta_{\rm gs}\big)$ are parameters specific to the storage platform: $\zeta$ is the ratio of the control-induced couplings from $\ket{s} \rightarrow \ket{e}$ (desired) and $\ket{g} \rightarrow \ket{e}$ (unwanted); $\gamma_{\rm eg}$ is the optical decoherence of the excited state ; and $\Delta_{\rm gs}$ is the hyperfine splitting between the $\ket{g}$ and $\ket{s}$ spin levels. Since the \big($d , \Omega_{\rm C}$\big) values required to optimally realize a memory protocol are proportional to the memory bandwidth $B$ along with proportionality constants specific to the dynamics of that protocol, $S_{\rm FWM}$ depends very strongly on the memory bandwidth and the resource scaling of the implementing protocol. Figure~\ref{fig:FWM} shows a comparison of the FWM noise strength among the SR, ATS, and EIT protocols for optimal broadband storage in a homogeneously-broadened $^{87}$Rb ensemble where $\gamma_{\rm eg}= \Gamma_{\rm eg}/2 = 2\pi\times$ 3 MHz, $\Delta_{\rm gs} = 2\pi \times$ 6.83 GHz, and $\zeta =$ 1.33. We calculate $S_{\rm FWM}$ using equation~(\ref{eq:FWM}) for bandwidths $10\Gamma_{\rm eg}/2\pi \le B \le 60\Gamma_{\rm eg}/2\pi$ corresponding to exponential probes of duration 0.4 ns $\le T_{\rm P} \le$ 2.6 ns. For each protocol, the optimal values of the peak optical-depth and peak control Rabi frequency associated with a given bandwidth are determined from the non-adiabatic / adiabatic operation conditions of that protocol (Fig.~3, main text). By combining those $d$ vs. $B$ and $\Omega_{\rm C}$ vs. $B$ scaling in equation~(\ref{eq:FWM}) for the range of bandwidths under consideration, we find that the probability of FWM noise affecting the SR memory is 3-4 orders of magnitude higher than the ATS memory (due to highest control power requirement) while it is still a factor of 3-7 times lower than the EIT memory (due to lowest optical-depth requirement) i.e., $S_{\rm FWM}^{\rm ATS} \ll S_{\rm FWM}^{\rm SR} < S_{\rm FWM}^{\rm EIT}$.

\section{Discerning storage based on SR and ATS protocols in non-optimal situations} \label{sec:SR-ATS}

\begin{figure}
\begin{center}
\includegraphics[width = 85 mm]{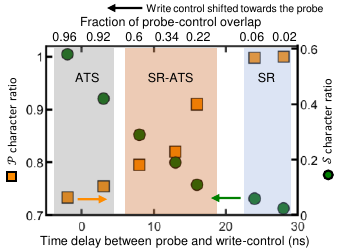}
\caption{Non-optimal SR implementation showing transition in the storage character from pure SR to mixed SR+ATS to ATS in our experiment. Different storage regimes are achieved by simply changing the arrival time of the write control with respect to the input probe.} 
\label{fig:SR-ATS}
\end{center}
\end{figure}

Storage in both SR and ATS protocols relies on the excitation of the atomic polarization coherence ($P(z,t)$), but with one subtle difference: in the ATS memory~\cite{saglamyurek2018coherent,PhysRevA.100.012314} the writing stage happens over the entire duration of the probe, in which the control field mediates both probe absorption (which generates $P(z,t)$) as well as the conversion from $P \rightarrow S$ at the same time. In contrast, the SR protocol has absorption and writing in two separate stages: the write control is applied only after the $P$ has maximised and conversion into $S$ happens only during the time the write control is on, which can be much shorter than the probe duration for optimal performance (Fig.~1, main text). This fundamental distinction leads to very different requirements for optimal implementation of the two memories such that the associated storage processes can never mix. However, in non-optimal SR implementations with control duration longer than the system's superradiant decay: if the absorption and writing are not well-separated such that there is a partial or complete overlap between the probe and write control fields, the storage maybe contributed by processes related to SR and ATS protocols both. Larger the probe-control overlap, greater is the ATS contribution to this `mixed' storage. To identify implementations where storage is mainly dominated by SR protocol's mechanism, especially in a non-optimal setting, we define the following dimensionless character factors:
\begin{enumerate}
    \item $\mathcal{P}$ is the ratio of the polarization build-up in the presence of write control to that in its absence.~In either case, the polarization build-up is computed over the entire probe duration ($\tau_{\rm p}$) and across the entire medium length ($L$) as $\frac{1}{L \tau_{\rm p}}\int_{\rm 0}^{\rm L}\int_{\rm 0}^{\rm \tau_{\rm p}}|P(z,t)|^{\rm 2} dz~ dt$, where $\tau_{\rm p} = 3.(T_{\rm P} + T_{\rm F})$ for an exponential probe.~Implementations for which $\mathcal{P} \rightarrow$ 1 correspond to pure SR storage with no or negligible probe-control overlap such that the write control does not affect the generated polarization. As the probe-control overlap increases, the value of $\mathcal{P}$ decreases due to its simultaneous mapping into the spin-wave, affecting storage via ATS protocol's mechanism. 
    
    \item $\mathcal{S}$ is the ratio of the spin-wave generated during the probe-control overlap time ($\Delta t$) to the total spinwave generated during writing ($\tau_{\rm c}$). The overlap spinwave can be computed using $\frac{1}{L \Delta t}\int_{\rm 0}^{\rm L}\int_{\rm 0}^{\Delta t}|S(z,t)|^{\rm 2} dz~ dt$ while the total spinwave as $\frac{1}{L \tau_{\rm c}}\int_{\rm 0}^{\rm L}\int_{\rm 0}^{\tau_{\rm c}}|S(z,t)|^{\rm 2} dz~ dt$, where for a Gauusian control with FWHM duration as $T_{\rm C}$, $\tau_{\rm c} =$ 2.25 $T_{\rm C}$. Memory implementations featuring $\mathcal{S} \rightarrow$ 0 represent pure SR storage in that no or negligible spin-wave is generated while the probe field is being absorbed.
\end{enumerate}
In figure~\ref{fig:SR-ATS}, we plot $\mathcal{P}$ and $\mathcal{S}$ character values for several memory implementations in our experiment, each associated with a storage mechanism that is based on either pure SR, or mixed SR + ATS, or pure ATS. The only parameter that we vary is the arrival-time of the write control after the input probe is applied, which we define as the time delay between their peak intensities. Other parameters are kept fixed and include: exponential probe with $T_{\rm P} =$ 10 ns and $B =$ 12.7 MHz; Gaussian control with $T_{\rm C} =$ 20 ns and $A_{\rm C}=$ 0.85$\pi$; $d =$ 9; $\gamma = 2\pi.$3 MHz. Larger delay times correspond to a small probe-control overlap such that the spin-wave storage is achieved only from the mechanism of the SR protocol~\big($\mathcal{P} \rightarrow$ 1, $\mathcal{S} \rightarrow$ 0\big). By shifting the write pulse closer to probe i.e., decreasing the arrival-time and increasing the overlap, storage is affected by the mechanisms of both ATS and SR protocols as evident from the decrease (increase) in $\mathcal{P}$ ($\mathcal{S}$). Finally, the ATS mechanism dominates when the write-control and probe are on almost at the same time and the overlap is close to maximum. Our results indicate that under non-optimized control setting, the transition from one memory to the other is smooth and can occur by a simple change of the control field's position relative to the probe. We also measure the experimental memory efficiency $\eta^{\rm exp}$ in each storage regime. Importantly, the (unoptimal) memory efficiency in the mixed regime can be higher than that in the true SR regime: for example, with 25 ns separation from the probe, the non-optimized control fields in our experiment yield an efficiency of 3\% in the true SR regime; shifting the write-control towards the probe by only a few nanoseconds can easily double this efficiency, however, the storage mechanism does not remain pure-SR based. By continually shifting the write-control in the pursuit of higher memory efficiency, one can enter the regime of pure yet unoptimal `ATS'-based storage yielding efficiency upto 10\%. We emphasize that in all cases, the efficiency remains less than what can be achieved using the optimal implementation of either protocol (having completely different probe-and-control requirements). This further stresses the need to operate a memory protocol in its native and optimal regime where the advantages associated with that protocol's mechanism are best met.


\begin{thebibliography}{68}
\bibitem [{Foo()}]{Footnote1}%
  \BibitemOpen
  \href@noop {} {\ }\bibinfo {note} {The term ``superradiance" may refer both
  to the case of externally imposed dipoles (as in this work) and the case of
  what is also commonly referred to as ``superfluorescence,'' in which dipoles
  self-align after a spontaneous emission event. The emission properties in
  both cases are the same.}\BibitemShut {Stop}%
\end{thebibliography}
\end{document}